\documentclass[manuscript]{aastex}
\usepackage{txfonts}
\usepackage{latexsym,bm}
\usepackage{lineno}
\usepackage{multirow}
\shorttitle{Spectra Invariance of Gradual SEPs}
\shortauthors{Wang and Qin }

\begin{document}

\title{SIMULATIONS OF THE SPATIAL AND TEMPORAL INVARIANCE IN THE SPECTRA OF GRADUAL 
SOLAR ENERGETIC PARTICLE EVENTS}

\author{ Yang Wang\altaffilmark{1} and Gang Qin\altaffilmark{1}}
\email{ywang@spaceweather.ac.cn; gqin@spaceweather.ac.cn}
\altaffiltext{1}{State Key Laboratory of Space Weather,  Center for Space Science and Applied Research, Chinese Academy of Sciences, Beijing 100190, China}

\begin{abstract}
The spatial and temporal invariance in the spectra of energetic particles in the gradual solar events is reproduced in the simulations.  
Based on a numerical solution of the focused transport equation, we obtain the intensity time profiles of  solar energetic particles (SEPs) accelerated by an interplanetary shock in the three-dimensional interplanetary space. 
The shock is treated as a moving source of energetic particles with a distribution function. 
The time profiles of particle flux with different energies  are calculated in the ecliptic at $1$ AU.  
According to our model, we find that shock acceleration strength, parallel diffusion and adiabatic cooling are the main factors in forming the  spatial invariance in SEP spectra, and perpendicular diffusion is a secondary factor.
In addition, the temporal invariance in SEP spectra is mainly due to the effect of adiabatic cooling.
Furthermore, a spectra invariant region, which agrees with observations but is different than the one suggested by Reames and co-workers, is proposed based on our simulations.

\end{abstract}


\section{INTRODUCTION}
Solar energetic particle (SEP) events can roughly be divided into two categories: impulsive events and gradual events \citep{Reames1995RvGeS..33..585R,Reames1999SSRv90413R}. 
The impulsive events, with the characteristics of low intensity and short duration, are produced by solar flares. 
Gradual events, usually lasting longer and having high intensity, are related to the shocks driven by interplanetary coronal mass ejections (ICMEs). 
\cite{Lario2006ApJ...653.1531L} investigated the radial and longitudinal dependence of $4-13$ and $27-37$ MeV proton peak intensities and fluences measured within $1$ AU.
They found the peak intensities and fluences of SEP events can be approximated by $j = {j_0}{r^{ - \alpha }}\exp \left[ { - k{{\left( {\phi  - {\phi _0}} \right)}^2}} \right]$, where $j$ is either the peak intensity or the fluence, $r$ is the radial distance of the spacecraft, $\phi$ is the longitudinal angular distance between the footpoint of the observer's field line and the region of the SEP source, and $\phi_0$ is the centroid of the distributions.
Furthermore, the radial dependence of peak intensities and fluences of SEP events have been simulated  with a focused-diffusion transport equation \citep{Lario2007AdSpR..40..289L}.

Generally, there are two major approaches to modeling SEP 
acceleration by CME driven shocks: some authors \citep{Heras1992ApJ...391..359H,Heras1995ApJ...445..497H, Kallenrode1997JGR...10222311K,Lario1998ApJ...509..415L,
Kallenrode2001JGR...10624989K,Ng2003ApJ...591..461N,wang2012effects,
qin2013transport} adopted a ``black box'' model to treat the shock as a moving source, and  SEPs are  injected at the shock with an assumed injection strength, while a few other studies include the acceleration of SEPs by shocks \citep{Lee1983JGR....88.6109L,Gordon1999JGR...10428263G, Zank2000JGR...10525079Z, Li2003JGRA..108.1082L, Rice2003JGRA..108.1369R,Sokolov2004ApJ...616L.171S, Li2005JGRA..110.6104L, kota2005AIPC..781..201K, Tylka2006ApJ...646.1319T,zuo2011, zuo2013acceleration}.
In these models, three important effects of acceleration and propagation mechanisms have been involved.
The first effect is the acceleration process by the CME-driven shock. \cite{Zank2000JGR...10525079Z}  modeled the evolution of a CME-driven shock based on an ``onion shell'' model, and this model has been furtherly developed in a number of papers
\citep{Li2003JGRA..108.1082L,Rice2003JGRA..108.1369R, Li2005JGRA..110.6104L}. 
They used a magnetohydrodynamics (MHD) code to describe the evolution of the CME-driven shock in the interplanetary space, and wave excitation by streaming energetic particles produced at shock is included. 
Based on the model, the simulation results can successfully explain the SEP fluxes and spectra in some multi-spacecraft observed events \citep{Verkhoglyadova2009ApJ...693..894V,Verkhoglyadova2010JGRA..11512103V}.
The second effect is energetic particles to interact with Alf\'ven wave self-consistently.
\cite{Ng2003ApJ...591..461N,Ng2012AIPC.1436..212N} presented a model of particle transport including streaming proton-generated Alf\'ven waves, and the amplification of the  Alf\'ven waves is  determined by the anisotropy of particles.
The particle diffusion coefficients can be calculated from  wave intensity and wave growth rates. Their simulation results show a good agreement with the observed spectral slope and abundance ratios of heavy ions.
The third effect is the realistic geometry of CME and its shock \citep{Sokolov2004ApJ...616L.171S, kota2005AIPC..781..201K}. 
\cite {Sokolov2004ApJ...616L.171S} modeled particle acceleration and transport as CME driven shock wave propagating from $4$ to $30$ solar radius from the Sun. 
The realistic structures of CME and its shock are derived from a numerical solution of a fully three-dimensional MHD model. Their simulation results demonstrate that the diffusive shock acceleration theory can account for the increase of hundreds of MeV protons during  the early stages of CME driven shock.

Solar energetic particle events measured by multi-spacecraft help us to understand the processes of particle acceleration and transport in the heliosphere.
In some gradual events, the SEP fluxes measured by widely separated spacecraft  present similar intensities within a small $ \sim 2-3$ factor in different latitudes, longitudes or radius \citep{Reames1997ApJ...491..414R,Reames2010SoPh..265..187R,Reames2013SSRv..175...53R,
McKibben2001ICRC....8.3281M,Maclennan2001AGUSM..SH31A05M,Lario2003AdSpR..32..579L,Tan2009ApJ...701.1753T}.
This phenomenon was firstly proposed by \cite{McKibben1972JGR....77.3957M}, and was named ``reservoir'' by
\cite{Roelof1992GeoRL..19.1243R}. 
In order to interpret the reservoir phenomenon, \cite{McKibben1972JGR....77.3957M} and \cite{McKibben2001ICRC....8.3281M} involved an effective perpendicular diffusion to reduce the spatial gradients of flux, while \cite{Roelof1992GeoRL..19.1243R} suggested a diffusion barrier produced by ICMEs or shocks. 
The magnitude of magnetic field increases at the outer boundary of reservoirs, so  that SEPs could  be contained in the reservoirs for a long time.
Furthermore, the interplanetary magnetic field (IMF) has been disturbed by ICMEs, SEPs could be redistributed. 
\cite{Reames1996ApJ...466..473R}  shows that, in some gradual SEP events, the spectra are invariant both in space and time. This discovery extended the original work of \cite{McKibben1972JGR....77.3957M}. 
In \cite{Reames1996ApJ...466..473R}, they considered an expanding  magnetic bottle of quasi-trapped particles between an ICME driven shock and the Sun. 
As the magnetic bottle expanding, the SEP fluxes gradually decrease as a result of parallel diffusion and adiabatic cooling. 
In this sense, the magnetic bottle plays  a pivotal role  in the decay phase of SEP event.

In principle, the disturbances in the magnetic field caused by ICMEs can help the particles redistribute in space.
However, the reservoir phenomenon cannot be simply explained as a result of the disturbances of IMF caused by ICMEs.
Firstly, in the redistribution process in \cite{Reames1996ApJ...466..473R} and \cite{Reames2013SSRv..175...53R}, no explicit transport mechanism can reduce latitudinal, longitudinal, and radial gradients of SEP fluxes besides perpendicular diffusion.   
Secondly,  in some SEP events, ICMEs are not directly observed by the spacecraft, the reservoir phenomenon is also observed 
\citep{McKibben2001SSRv}. 
Thirdly, in  \cite{Reames1999SSRv90413R}, when the observer is located at the eastern of the shock, the onset time of temporal invariance in the SEP spectra is earlier than the shock arrival time. 
These results are not consistent with that of an expanding magnetic bottle.

The effect of perpendicular diffusion is important in the SEP fluxes especially when the observer is disconnected from the shock by IMF.  
During the time period March $1$, $1979$ to March $11$, $1979$, a gradual SEP event has been detected by $Helios$ $1$, $Helios$ $2$, and $IMP$ $8$. 
The three spacecraft are located  in the ecliptic near $1$ AU, but at different longitudes.
In the decay phase of this SEP event, the reservoir phenomenon appeared \citep{Reames1997ApJ...491..414R,Reames1999SSRv90413R,Reames2010SoPh..265..187R,Reames2013SSRv..175...53R}.
In this event, the in-situ observation shows that an ICME was detected by $Helios$ $1$, but not by $Helios$ $2$ and $IMP$ $8$. 
And the interplanetary shock was only observed by $Helios$ 1 and 2, but not by $IMP$ $8$ \citep{Lario2006ApJ...653.1531L,Reames2010SoPh..265..187R}. 
According to the location of the three spacecraft,  if the ICME was located behind the center of the shock front, then $Helios$ $1$ was located near to the center of shock, and $Helios$ $2$ and $IMP$ $8$ were located at the West frank of the shock.
However, the onset time of SEP fluxes observed by three spacecraft was very close. 
How could SEPs be detected by $IMP$ $8$ before it was connected to the shock by IMF? 
One possible answer is the effect of perpendicular diffusion which also possibly works in forming the reservoir phenomenon.


However, perpendicular diffusion has been always a difficult problem for several decades. 
Observation results show various levels of perpendicular diffusion coefficients for different SEP events.
For example, `dropout' phenomenon in the impulsive SEP event \citep{mazur2000ApJInterplanetary} usually show reduced perpendicular diffusion, in order to reproduce the `dropout' phenomenon in simulations, perpendicular diffusion coefficient ${{{\kappa }}_ \bot }$ should be several order magnitude smaller than  parallel  one ${{{\kappa }}_\parallel } $ \citep{giacalone2000ApJLSmall, guo2013ApJSmall,droge2010ApJ, Wang2014ApJ789157W}. 
On the other hand, for some events, observation results show that the perpendicular diffusion coefficients could be comparable to the parallel ones \citep{Dwyer1997ApJ...490L.115D,Zhang2003ApJ...595..493Z, Dresing2012SoPh}.
In order to understand diffusion, many efforts were made theoretically. 
By assuming energetic particles' perpendicular and parallel diffusion do not interaction, \citet{Jokipii1966ApJ...146..480J} developed the quasi-linear theory (QLT). 
According to QLT, perpendicular diffusion coefficient is usually much smaller than the parallel one. 
However, it is found that interaction between parallel and perpendicular diffusion is important in theory \citep{Kota2000ApJVelocity} and in simulations \citep{Qin2002GeoRLSubdiffusive, Qin2002ApJ...578L.117Q}, so the non-linear guiding center (NLGC) theory \citep{Matthaeus2003ApJ...590L..53M} is developed to describe perpendicular diffusion with the influence of parallel diffusion, which agrees with simulations much better than QLT. In addition, simulations show different levels of perpendicular diffusion, e.g., in \cite{Qin2012AdSpRNumerical}, the magnitude of ${{{\kappa }}_ \bot }/{{{\kappa }}_\parallel }$ could be as large as $10^{-1}$ in some conditions, and as small as $10^{-4}$ in other conditions.

Recently, \cite{qin2013transport} proposed that the shock acceleration strength makes important contributions to the reservoir phenomenon, particularly in low-energy SEPs. 
In their simulations, the reservoir phenomenon is reproduced under a variety of conditions of shock acceleration strength and perpendicular diffusion.
In this paper, as a continuation of \citet{qin2013transport}, we study the property of SEP spectra in the decay phase. 
We compute the time profiles of SEP flux which are accelerated by interplanetary shock. In section 2 we describe the SEP transport model and the shock model. 
In Section 3 we show the simulation results. 
In Section 4  we summary our results.

\section{MODEL}
In this work, we model the transport of SEPs following previous research\citep[e.g.,][]{Qin2006JGRA..11108101Q,Zhang2009ApJ...692..109Z,droge2010ApJ,he2011propagation, zuo2011,wang2012effects,qin2013transport,zuo2013acceleration,Wang2014ApJ789157W}.
A three-dimensional focused transport equation is written as \citep{Skilling1971ApJ...170..265S,schlickeiser2002cosmic,Qin2006JGRA..11108101Q,Zhang2009ApJ...692..109Z}
\begin{eqnarray}
  \frac{{\partial f}}{{\partial t}} = \nabla\cdot\left( \bm
  {\kappa_\bot}
\cdot\nabla f\right)- \left(v\mu \bm{\mathop b\limits^ \wedge}
+ \bm{V}^{sw}\right)
\cdot \nabla f + \frac{\partial }{{\partial \mu }}\left(D_
{\mu \mu }
\frac{{\partial f}}{{\partial \mu }}\right) \nonumber \\
  + p\left[ {\frac{{1 - \mu ^2 }}{2}\left( {\nabla  \cdot \bm{V}^
  {sw}  -
\bm{\mathop b\limits^ \wedge  \mathop b\limits^ \wedge } :\nabla
\bm{V}^{sw} } \right) +
\mu ^2 \bm{\mathop b\limits^ \wedge  \mathop b\limits^ \wedge}  :
\nabla \bm{V}^{sw} }
\right]\frac{{\partial f}}{{\partial p}} \nonumber \\
  - \frac{{1 - \mu ^2 }}{2}\left[ { - \frac{v}{L} + \mu \left
  ( {\nabla  \cdot
\bm{V}^{sw}  - 3\bm{\mathop b\limits^ \wedge  \mathop b\limits^
\wedge}  :\nabla \bm{V}^{sw} }
\right)} \right]\frac{{\partial f}}{{\partial \mu }},\label{dfdt}
\end{eqnarray}
where $f(\bm{x},\mu,p,t)$ is the gyrophase-averaged distribution function; $\bm{x}$ is the position in a non-rotating heliographic coordinate system; $t$ is the time;  $\mu$, $p$, and $v$ are the  particle pitch-angle cosine, momentum, and speed, respectively, in the solar wind frame; $\bm{\mathop b\limits^ \wedge}$ is a unit vector along the local magnetic field; $\bm{V}^{sw}=V^{sw}\bm{\mathop r\limits^ \wedge}$ is the solar wind velocity;  and $L$ is the magnetic focusing length given by $L=\left(\bm{\mathop b\limits^ \wedge}\cdot\nabla\\{ln} B_0\right)^{-1}$ with $B_0$ being the magnitude of the background  magnetic field. 
The IMF is set as the Parker field model, and the solar wind speed is  $400$ km/s. 
This equation includes many important particle transport effects such as particle streaming along the field line,  adiabatic cooling in the expanding solar wind, magnetic focusing in the diverging IMF, and the diffusion coefficients parallel and perpendicular to the IMF.

The pitch angle diffusion coefficient model is set as  \citep{Beeck1986ApJ...311..437B,qin2005model}
\begin{equation}
 D_{\mu \mu }  = D_0 \upsilon p^{ q-2} \left\{ {\left. {\left| \mu  \right.} \right|^{q - 1}  + h}
\right\}\left( {1 - \mu ^2 } \right),\label{D_mu_mu}
\end{equation}
where the constant $D_0$ controls the magnetic field fluctuation level. The constant $q$ is chosen  as $5/3$ for a Kolmogorov spectrum type of the power spectra density of magnetic field turbulence in the inertial range.
Furthermore, $h=0.01$ is chosen for the non-linear effect of pitch-angle diffusion at $\mu=0$ in the solar wind \citep{qin2009pitch,qin2014detailed}.

The parallel mean free path (MFP) $\lambda _\parallel$ can be written as \citep{Jokipii1966ApJ...146..480J,hasselmann1968scattering,earl1974diffusive}
\begin{equation}
\lambda _\parallel   = \frac{{3\upsilon}}{8}\int_{ - 1}^{ + 1}
{\frac{{(1 - \mu ^2 )^2 }}{{D_{\mu \mu } }}d\mu
},\label{lambda_parallel_1}
\end{equation}
and the parallel diffusion coefficient $\kappa_\parallel$ can be written as $\kappa_\parallel=v\lambda_\parallel/3$.

The relation of the particle momentum  and the perpendicular diffusion coefficient is set as \citep{Potgieter1985ApJ...294..425P,Zhang1999ApJ...513..409Z}
\begin{equation}
{{\bm{\kappa }}_ \bot } = {\kappa _0}\left( {\frac{v}{c}} \right){\left( {\frac{p}{{1{\kern 3pt} {\rm{GeV}}{c^{ - 1}}}}} \right)^{\alpha}}\left( {\frac{{{B_e}}}{B}} \right)\left( {{\bf{I}} - \mathop {\bf{b}}\limits^ \wedge  \mathop {\bf{b}}\limits^ \wedge  } \right)
\end{equation}
where $B_e$ is magnetic field strength at the Earth, $B$ is the magnetic field strength at the location of particle, $p$ is  particle momentum, and $\alpha$ is set to $1/3$. Different perpendicular diffusion coefficients could be obtained by altering $\kappa _0$.
Note that we use this ad-hoc model for the purpose of simplicity, the parameters, e.g., $\alpha$ could be set as other values \citep{Zhang1999ApJ...513..409Z}. 
However, the variation of these parameters would not qualitatively change the results in this paper.
There are some more complete models that are developed to describe the particle diffusion in magnetic turbulence, such as the nonlinear guiding center theory 
\citep{Matthaeus2003ApJ...590L..53M,Shalchi2004ApJ...616..617S,Shalchi2010Ap&SS.325...99S,qin2014modification}.

We use a time-backward Markov stochastic process method to solve the transport equation (\ref{dfdt}). The detail of method can be found in \cite{Zhang1999ApJ...513..409Z} and \cite{Qin2006JGRA..11108101Q}. 
The particle injection on the shock is specified by boundary values.
The boundary condition is chosen as following form \citep{Kallenrode1997JGR...10222311K,Kallenrode2001JGR...10624989K,wang2012effects,
qin2013transport}
\begin{eqnarray}
 f_b (r,\theta ,\varphi ,p,t) &=& a \cdot \delta (r - \upsilon_s t) \cdot S(r,\theta ,\varphi ,p)
 \cdot p^{ - \gamma }  \cdot \xi (\theta ,\varphi ) \nonumber\\
 S(r,\theta ,\varphi ,p) &=& (\frac{r}{{r_c }})^{-\alpha (p)}  \cdot \exp \left[ { - \frac{{\left|
 {\phi (\theta ,\varphi )} \right|}}{{\phi _c (p)}}} \right] \nonumber\\
 \xi (\theta ,\varphi ) &=& \left\{ \begin{array}{l}
 1{\kern 21pt}  \rm{if} \left| {\phi (\theta ,\varphi )}
 \right| \le \phi _s  \\
 0{\kern 21pt}  {\rm{otherwise,}} \\
 \end{array} \right.\label{f_b}
\end{eqnarray}
where the particle are injected at $r = \upsilon_s t$, $\upsilon_s$
is shock speed. $\upsilon_s t=r_0 + n \cdot \Delta r$, with $n =
0,1,2 \cdot \cdot \cdot n_0$. $\Delta r$ is space interval between
two `fresh' injections, $r_0 = 0.05$ $AU$ is inner boundary. 
$r_c$ is set to $0.05$. $r$ is distance between sun and shock. 
${\phi}$ is the angle between the center of shock and the point at  the shock front where the particles injected.
The shock acceleration strength is  set as $S$ for specifying the particles ejection.
It changes with a power law in radial distance and exponential towards the flank of shock.
$\xi$ determines the spatial scale of shock front. 
${\phi_s}$ is the half width of the shock.

\section{RESULTS}

The parameters used are listed in table \ref{paratable}, unless otherwise stated. Note that the IMF is set as Parker spiral, and the disturbances of IMF behind the shock are ignored. 
The particle energy channels are chosen as  $5$ MeV, $10$ MeV, $20$ MeV, $40$ MeV and $80$ MeV. The parallel mean free path depends on the momentum ${\lambda _\parallel }\sim p^{1/3}$. 
According to \cite{qin2013transport}, the ${{{\kappa }}_ \bot }/{{{\kappa }}_\parallel }$ is set as $0.1$ in the ecliptic at $1$ AU.
Because the shock acceleration efficient decreases as the particle energy increasing, the acceleration strength parameters also change with the momentum: $\alpha \sim p^{0.3}$, $\phi_c \sim p^{-0.3}$. 
The observers are located in the ecliptic at $1$ AU.

\subsection{Temporal Invariance in the Spectra}

\begin{figure}
\epsscale{1} \plotone{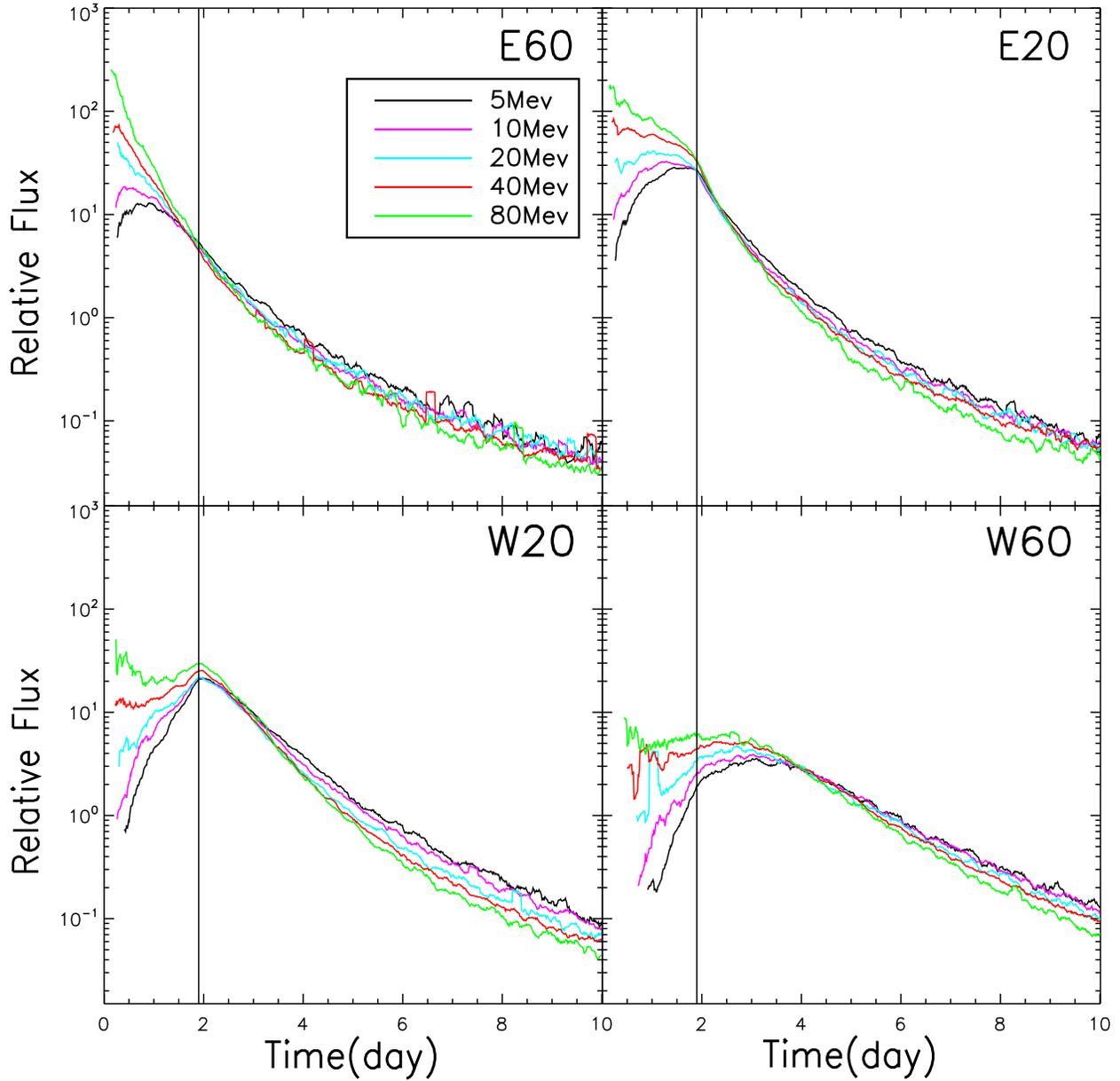} 
\caption{Comparison of different energy protons observed by observers in the $1$ AU ecliptic at different longitudes. 
The fluxes are normalized so their values are similar after all of them reach peaks. 
The vertical lines indicate the shocks passage of $1$ AU. 
The adiabatic cooling is included in simulations.
\label{time_invariant}}
\end{figure}

\begin{figure}
\epsscale{1} \plotone{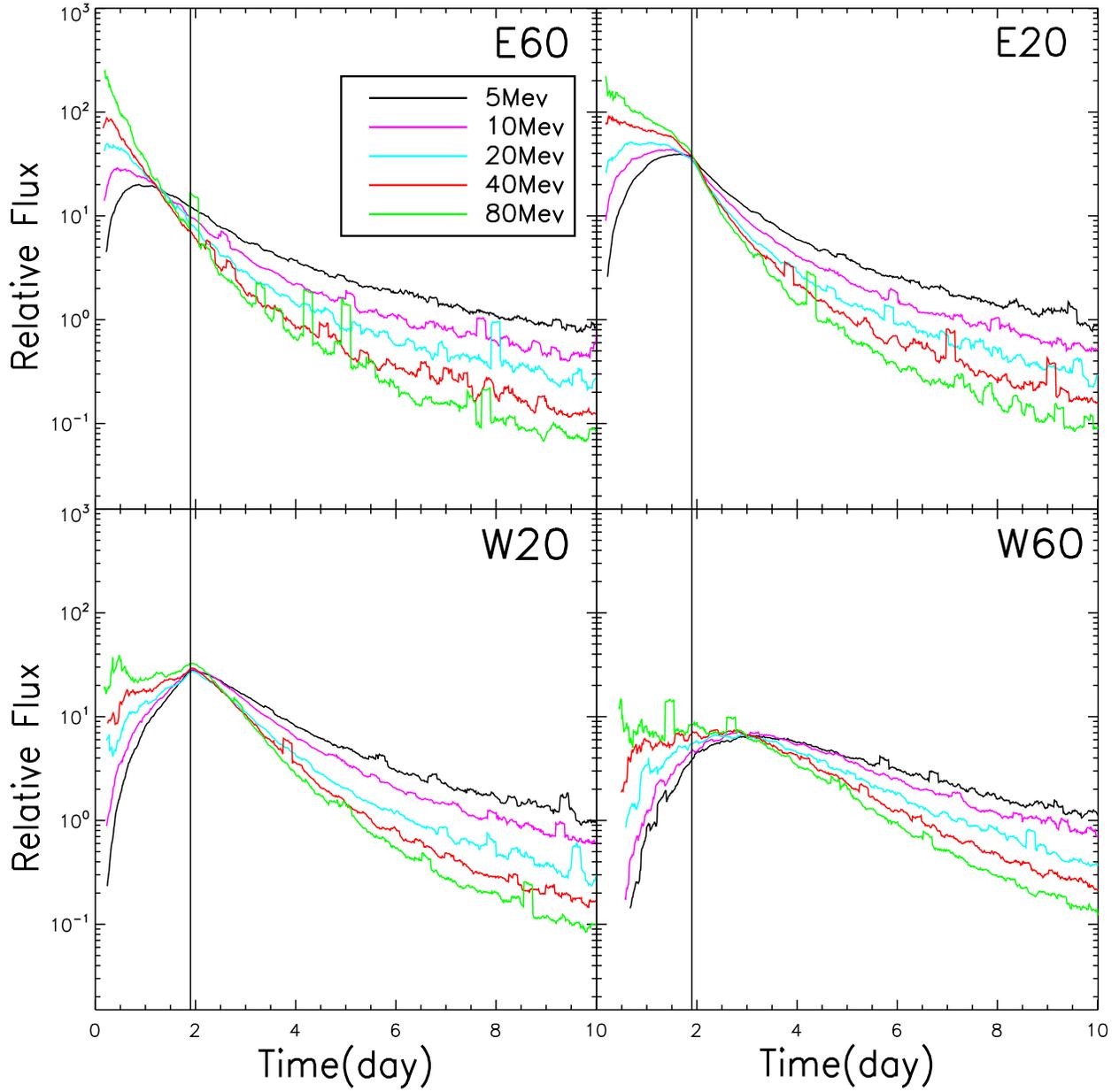} 
\caption{Same as \ref{time_invariant} except that the adiabatic
cooling is not included.
\label{without_cooling_dif_E}}
\end{figure}

In Figure \ref{time_invariant} and \ref{without_cooling_dif_E}, we plotted the fluxes of different energy channels in the cases with and without adiabatic cooling. 
In order to check the temporal properties of SEP spectra in the decay phase, we normalize different energy fluxes, so that the fluxes have similar values soon after all of them have reached peaks.
The panels show the normalized fluxes observed in the ecliptic at $1$ AU, but at different longitudes E$60$, E$20$, W$20$ and W$60$. 
The notations E$60$, E$20$, W$20$, W$60$ are short for East $60^\circ$, East $20^\circ$, West $20^\circ$ and West $60^\circ$, respectively. 
East/West means the location of the observer is east/west relative to the center of shock. 
The vertical lines indicate the shocks passage of the observers.

In Figure \ref{time_invariant}, the adiabatic cooling  effect is included in SEP propagation process. 
In the decay phase of SEP events, shock acceleration strength,  adiabatic cooling, parallel diffusion, and perpendicular diffusion are the major factors to influence the flux behavior. 
In the four panels, the fluxes at all energies follow a similar trend, then the fluxes scatter slowly as time goes by. 
This is called temporal invariance in the spectra of gradual SEP event. 
In the E$60$ event, the fluxes at all energies start to follow a similar trend about one day before the shock passage of $1$ AU. 
In other words, the onset time of the temporal invariance is earlier than the time of the shock passage of the observer. 
In the E$20$ and W$20$ events, however, the onset time of temporal invariance is close to the shock passage of the observers. 
In the W$60$ event, furthermore, the temporal invariance starts the latest, and actually it starts two days latter than the shock arrival.

In Figure \ref{without_cooling_dif_E}, the adiabatic cooling is not included in the SEP propagation process. 
Without adiabatic cooling, shock acceleration strength, parallel and perpendicular diffusion are the major factors in the decay phase. 
Due to the different diffusion coefficients and shock acceleration strength for different energy particles, the fluxes decay with different ratios consequently. 
With higher energies, the fluxes decay much faster. 
In these cases, the temporal invariance does not exist in the decay phase.

Comparing Figure \ref{time_invariant} and Figure \ref{without_cooling_dif_E}, the fluxes decrease much faster with adiabatic cooling. 
Because of adiabatic energy loss, particles have less energy when they are observed than that when they are released in the sources. 
In addition, since the source spectrum index is negative, the fluxes are lower with higher energies, so that the adiabatic cooling effect makes the SEP flux decreasing as time passes by. 
To sum up, the temporal invariance in the spectra results from the adiabatic cooling effect.

\subsection{Spatial Invariance in the Spectra}
\begin{figure}
\epsscale{1} \plotone{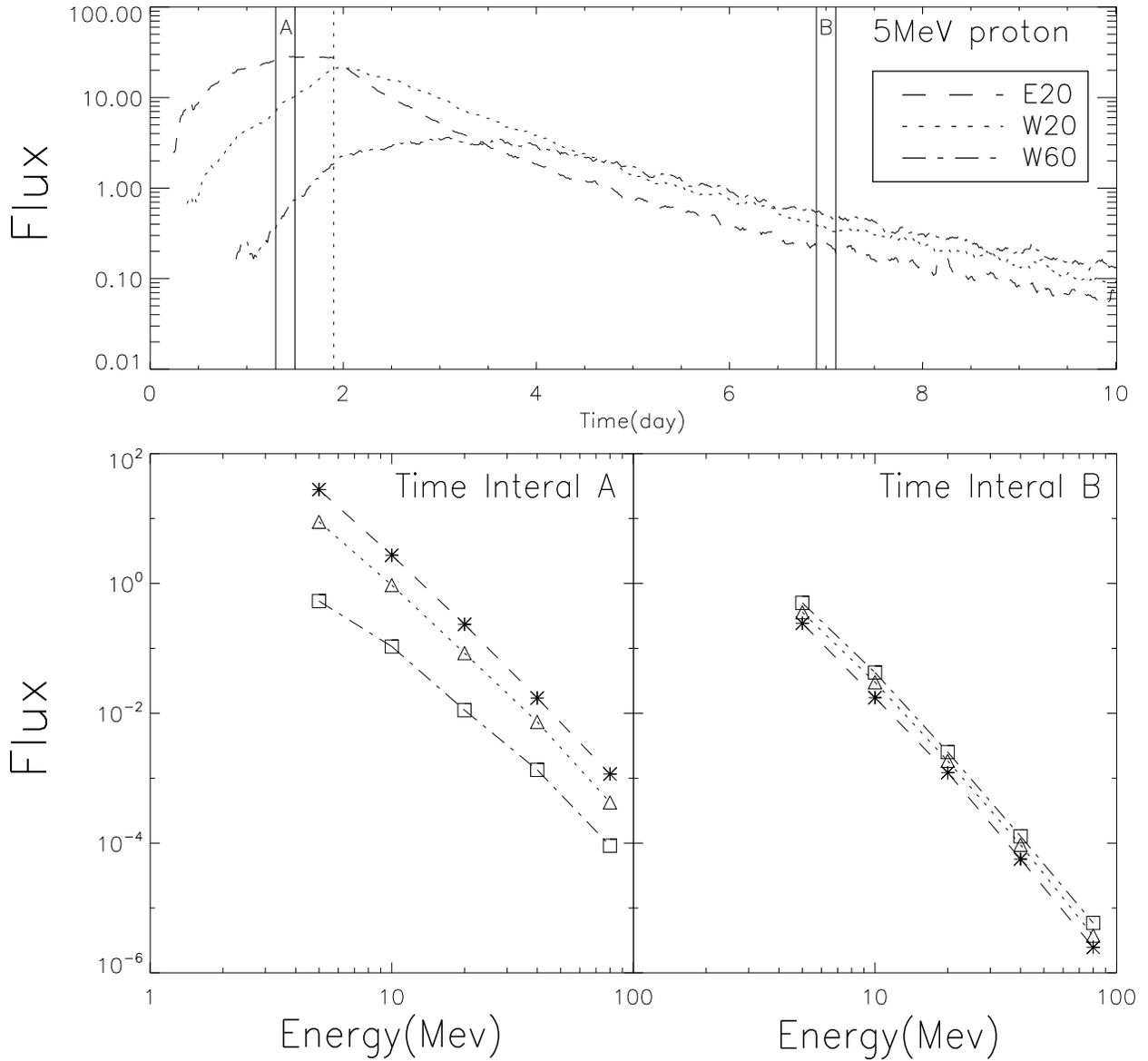} 
\caption{In the upper panel, comparison of $5$ MeV protons flux observed by the observers in $1$ AU ecliptic at different longitudes. 
The lower left and right panels show the spectra observed at different longitudes during time interval $A$ and $B$, respectively. 
The vertical dashed lines indicate the shock passage of $1$ AU. 
\label{spatial_invariant}}
\end{figure}

In Figure \ref{spatial_invariant}, the SEP fluxes are shown for three observers located at different longitudes, E$20$, W$20$, and W$60$.  
The upper panel shows the $5$ MeV proton fluxes observed by the observers. We set two typical time intervals, interval $A$ from $1.3$ days to $1.5$ days in rising phase and interval $B$ from $6.9$ days to $7.1$ days in decay phase. 
In order to study the spatial variance in different phases, in lower left and right panels of Figure \ref{spatial_invariant}, we plot the energy spectra observed in different longitudes in interval $A$ and interval $B$, respectively.
During the interval $A$, the spectra are different among the three observers. However, during the interval $B$, spectra are almost the same among the three observers. 
This phenomenon, which is named spatial invariance in the spectra by \cite{Reames1997ApJ...491..414R}, results from the reservoir effect in different energy channels.

Because the shock is a moving source in the interplanetary space, the peak intensity of SEP flux is mainly determined by the shock acceleration strength and parallel MFP. 
In upper panel of Figure \ref{spatial_invariant}, at the peak time of flux for  W$20$ (W$60$),  the flux for W$20$ (W$60$) is close to that for E$20$.
Furthermore, the SEP fluxes decay in a similar ratio because of the effect of adiabatic cooling.  
At the same time,  the latitudinal gradient in the SEP fluxes is further reduced because of the effect of perpendicular diffusion.
However, in other simulations with different shock acceleration strength and  parallel MFP (not shown here), if at the peak time of flux for W$20$ (W$60$), the flux for W$20$ (W$60$) is significantly different than that for E$20$, the reservoir phenomenon can not form in normal diffusion coefficients.
As a result, shock acceleration strength, parallel diffusion,  and adiabatic cooling are the main factors in forming the reservoir phenomenon, and  perpendicular diffusion is a secondary factor.

\subsection{Invariant Spectra Region}
There are some important characteristics in the invariant spectra region from our 
simulations (Figure \ref{time_invariant}). If the observer is located at the 
eastern flank of the shock, the onset time of invariant spectra 
is earlier than the shock arrival. 
But if the observer is located near the central flank of shock, the spectra 
invariance begins approximately at the shock passages. Finally, if the observer is 
located at the western flank of shock, the onset time of invariant spectra are much 
later than the shock arrivals. From these results, we can better understand the 
invariant spectra region. 

\begin{figure}
\epsscale{0.6} \plotone{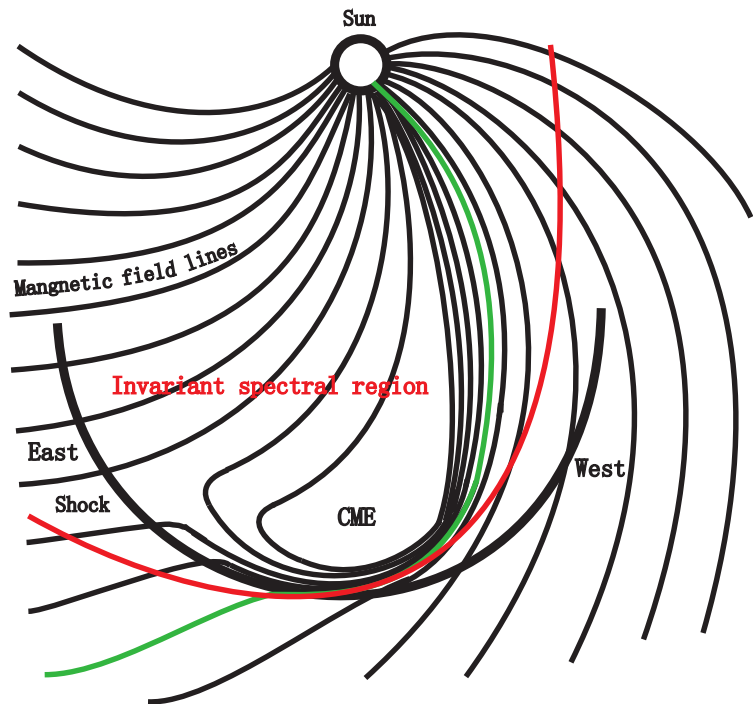} 
\caption{ The green line indicates the original spectra invariant region proposed by \cite{Reames1997ApJ...491..414R}. 
The red line indicates a new spectra invariant region based on the simulation results in this paper. 
\label{spectral_cartoon}}
\end{figure}

Figure \ref{spectral_cartoon} shows the invariant spectra region. In the picture, the green line is plotted by \cite{Reames1997ApJ...491..414R}, and the red line is new in this work. 
According to \cite{Reames1997ApJ...491..414R}, the left side of the green line is the invariant spectra region, with the assumption that particles are quasi-trapped in the region behind the ICME, and the SEP fluxes gradually decrease as a result of parallel diffusion and adiabatic deceleration mechanisms, and in addition, there are also some leakage of energetic particles from ICME to eastern side of the upstream shock. 
In this sense,  ICMEs play the pivotal role in the decay  phase of fluxes. 
As a result, the invariant spectra region is determined by ICMEs' propagation path plus some eastern side of upstream region. 
However, we suppose the invariant spectra region could be in the left side of the red line instead. 
In our simulations, ICME is not included, but in the propagation process perpendicular diffusion is included to reduce the spatial gradient in the fluxes, and adiabatic cooling is included to reduce the temporal variance. 
As the simulation results showed above, the spectra spatial and temporal invariance could result from  the effects of shock acceleration strength, adiabatic cooling, and perpendicular diffusion. 
In this sense, it is possible that the invariant spectra region is not confined by the ICMEs' propagation path. 
Instead, the invariant spectra region could be confined by the interplanetary shock, but the region expands faster (slower) than the shock at the eastern (western) flank, respectively.

\section{DISCUSSION AND CONCLUSIONS}
We have studied interplanetary shock accelerated SEPs propagation in three-dimensional IMF. The spectra observed by different observers are calculated, and the spatial and temporal invariance in the spectra are reproduced in the simulations. The following are our major findings.

The adiabatic cooling effect is the key factor for forming temporal invariance in the spectra. 
By including the adiabatic cooling, for different energy channels, the flux decay ratios are almost the same. 
The temporal invariance results from the fact that all energy particles decay as the same ratios because of adiabatic cooling effect. 
At the eastern flank of the shock, the onset time of the spectra invariance is earlier than the shock arrival. 
For the central cases, however, the onset time of the spectra invariance is close to the time of shock arrival. 
At the western flank of the shock, finally, the onset time of the spectra invariance is later than shock arrival.  
In addition, the fluxes decay much faster in the cases with adiabatic cooling. 
Without adiabatic cooling, the decay phase of SEP fluxes are dominated by shock acceleration strength, parallel diffusion, and perpendicular diffusion, which are all varying with particles' energies.  
Therefore, the temporal invariance does not exist without adiabatic cooling.

Shock acceleration strength, parallel diffusion, adiabatic cooling, and perpendicular diffusion are four important factors in forming the spatial invariance, which is in  reservoir phenomenon in different energy channels.
Shock acceleration strength parameters $\alpha$ and ${\phi _c}$ are set to $2$ and $15^\circ$ for $5$ MeV protons in our simulations, respectively. 
And these parameters also change with the momentum: $\alpha \sim p^{0.3}$, $\phi_c \sim p^{-0.3}$, because the shock acceleration strength decreases with higher energy particles. 
Among the four factors, shock acceleration strength,  parallel diffusion, and adiabatic cooling are the main factors in forming the reservoir phenomenon, and  perpendicular diffusion is a secondary one.
This conclusion  is derived based on our simulations, and it is also consistent with the observations. 
In \cite{Reames1997ApJ...491..414R} and \cite{Reames2013SSRv..175...53R},  a gradual SEP event was detected by $Helios$ $1$, $Helios$ $2$, and $IMP$ $8$ during  March $1$, $1979$ to March $11$, $1979$. 
In this event, the reservoir phenomenon appeared, and the onset time of SEPs observed by different spacecraft are very close because of the effect of perpendicular diffusion.
At the peak time of flux observed by $Helios$ 2 ($IMP$ 8),  the flux observed by $Helios$ 2  ($IMP$ 8) is close to that observed by $Helios$ 1.
The importance of the peak intensity of SEP flux observed  by $Helios$ 2  and $IMP$ 8 in forming the reservoir phenomenon  is also noticed by \cite{Reames2013SSRv..175...53R},
however, the reservoir phenomenon is explained as a result of the disturbance of IMF caused ICMEs.
In our model, the peak of flux is mainly determined by  shock acceleration strength and  parallel diffusion.
Furthermore, the SEP fluxes decay as a similar ratio because of the effect of adiabatic cooling.
At the same time,  the latitudinal gradient in the SEP fluxes is further reduced because of the effect of perpendicular diffusion.
Finally, according to our model, the reservoir phenomenon appeared in this SEP event with the effects of  shock acceleration strength,  parallel diffusion, adiabatic cooling, and perpendicular diffusion.
Observationally, shock acceleration strength, diffusion coefficients,  and adiabatic cooling change significantly in different SEP events  \citep{Kallenrode1996JGR...10124393K,Kallenrode1997JGR...10222347K}.
As a result, the reservoir phenomenon can only form in some gradual SEP events with those controlling effect parameters in appropriate values.

Based on our simulations, a new invariant region is proposed.
The new region is different from the one proposed by \cite{Reames1997ApJ...491..414R}. 
There are two important characteristics in our new region. 
First, if the observer is located at the eastern (western) flank of the shock, the onset time of temporal invariant in the spectra is earlier (later) than the shock arrival, respectively. 
Second, the spatial invariance in the spectra can also be formed without ICMEs. 
These two characteristics are supported by observations, but are difficult to be explained in the previous model \cite{Reames1997ApJ...491..414R}.

In our model, we ignore the disturbance of the IMF caused by ICME for the simplicity. 
In principle, the disturbance in the magnetic field can help particles redistribute in space. 
In future work, we intend to include a realistic three-dimensional ICME shock, so that the SEP acceleration and transport in the heliosphere can be investigated more precisely.

\acknowledgments
The authors thank the anonymous referee for valuable comments. We are partly supported by grants   NNSFC 41374177,  NNSFC 41125016, and NNSFC 41304135, the CMA grant GYHY201106011, and the Specialized Research Fund for State Key Laboratories of China. The computations were performed by Numerical Forecast Modeling R\&D and VR System of State Key Laboratory of Space Weather and Special HPC work stand of Chinese Meridian Project.

\bibliography{cankao.bib}

\begin{thebibliography}{69}
\expandafter\ifx\csname natexlab\endcsname\relax\def\natexlab#1{#1}\fi

\bibitem[{{Beeck} \& {Wibberenz}(1986)}]{Beeck1986ApJ...311..437B}
{Beeck}, J., \& {Wibberenz}, G. 1986, \apj, 311, 437

\bibitem[{{Dresing} {et~al.}(2012){Dresing}, {G{\'o}mez-Herrero}, {Klassen},
  {Heber}, {Kartavykh}, \& {Dr{\"o}ge}}]{Dresing2012SoPh}
{Dresing}, N., {G{\'o}mez-Herrero}, R., {Klassen}, A., {Heber}, B.,
  {Kartavykh}, Y., \& {Dr{\"o}ge}, W. 2012, \solphys, 281, 281

\bibitem[{{Dr{\"o}ge} {et~al.}(2010){Dr{\"o}ge}, {Kartavykh}, {Klecker}, \&
  {Kovaltsov}}]{droge2010ApJ}
{Dr{\"o}ge}, W., {Kartavykh}, Y.~Y., {Klecker}, B., \& {Kovaltsov}, G.~A. 2010,
  \apj, 709, 912

\bibitem[{{Dwyer} {et~al.}(1997){Dwyer}, {Mason}, {Mazur}, {Jokipii}, {von
  Rosenvinge}, \& {Lepping}}]{Dwyer1997ApJ...490L.115D}
{Dwyer}, J.~R., {Mason}, G.~M., {Mazur}, J.~E., {Jokipii}, J.~R., {von
  Rosenvinge}, T.~T., \& {Lepping}, R.~P. 1997, \apjl, 490, L115+

\bibitem[{Earl(1974)}]{earl1974diffusive}
Earl, J. 1974, The Astrophysical Journal, 193, 231

\bibitem[{Giacalone {et~al.}(2000)Giacalone, Jokipii, \&
  Mazur}]{giacalone2000ApJLSmall}
Giacalone, J., Jokipii, J., \& Mazur, J. 2000, The Astrophysical Journal
  Letters, 532, L75

\bibitem[{{Gordon} {et~al.}(1999){Gordon}, {Lee}, {M{\"o}bius}, \&
  {Trattner}}]{Gordon1999JGR...10428263G}
{Gordon}, B.~E., {Lee}, M.~A., {M{\"o}bius}, E., \& {Trattner}, K.~J. 1999,
  \jgr, 104, 28263

\bibitem[{Guo \& Giacalone(2014)}]{guo2013ApJSmall}
Guo, F., \& Giacalone, J. 2014, The Astrophysical Journal, 780, 16

\bibitem[{Hasselmann(1968)}]{hasselmann1968scattering}
Hasselmann, K. 1968, Z. Geophys., 34, 353

\bibitem[{He {et~al.}(2011)He, Qin, \& Zhang}]{he2011propagation}
He, H.-Q., Qin, G., \& Zhang, M. 2011, The Astrophysical Journal, 734, 74

\bibitem[{{Heras} {et~al.}(1995){Heras}, {Sanahuja}, {Lario}, {Smith},
  {Detman}, \& {Dryer}}]{Heras1995ApJ...445..497H}
{Heras}, A.~M., {Sanahuja}, B., {Lario}, D., {Smith}, Z.~K., {Detman}, T., \&
  {Dryer}, M. 1995, \apj, 445, 497

\bibitem[{{Heras} {et~al.}(1992){Heras}, {Sanahuja}, {Smith}, {Detman}, \&
  {Dryer}}]{Heras1992ApJ...391..359H}
{Heras}, A.~M., {Sanahuja}, B., {Smith}, Z.~K., {Detman}, T., \& {Dryer}, M.
  1992, \apj, 391, 359

\bibitem[{{Jokipii}(1966)}]{Jokipii1966ApJ...146..480J}
{Jokipii}, J.~R. 1966, \apj, 146, 480

\bibitem[{{Kallenrode}(1996)}]{Kallenrode1996JGR...10124393K}
{Kallenrode}, M. 1996, \jgr, 101, 24393

\bibitem[{{Kallenrode}(1997)}]{Kallenrode1997JGR...10222347K}
---. 1997, \jgr, 102, 22347

\bibitem[{{Kallenrode}(2001)}]{Kallenrode2001JGR...10624989K}
---. 2001, \jgr, 106, 24989

\bibitem[{{Kallenrode} \& {Wibberenz}(1997)}]{Kallenrode1997JGR...10222311K}
{Kallenrode}, M., \& {Wibberenz}, G. 1997, \jgr, 102, 22311

\bibitem[{{K{\'o}ta} \& {Jokipii}(2000)}]{Kota2000ApJVelocity}
{K{\'o}ta}, J., \& {Jokipii}, J.~R. 2000, \apj, 531, 1067

\bibitem[{{K{\'o}ta} {et~al.}(2005){K{\'o}ta}, {Manchester}, {Jokipii}, {de
  Zeeuw}, \& {Gombosi}}]{kota2005AIPC..781..201K}
{K{\'o}ta}, J., {Manchester}, W.~B., {Jokipii}, J.~R., {de Zeeuw}, D.~L., \&
  {Gombosi}, T.~I. 2005, in American Institute of Physics Conference Series,
  Vol. 781, The Physics of Collisionless Shocks: 4th Annual IGPP International
  Astrophysics Conference, ed. G.~{Li}, G.~P. {Zank}, \& C.~T. {Russell},
  201--206

\bibitem[{{Lario} {et~al.}(2007){Lario}, {Aran}, {Agueda}, \&
  {Sanahuja}}]{Lario2007AdSpR..40..289L}
{Lario}, D., {Aran}, A., {Agueda}, N., \& {Sanahuja}, B. 2007, Advances in
  Space Research, 40, 289

\bibitem[{{Lario} {et~al.}(2006){Lario}, {Kallenrode}, {Decker}, {Roelof},
  {Krimigis}, {Aran}, \& {Sanahuja}}]{Lario2006ApJ...653.1531L}
{Lario}, D., {Kallenrode}, M.-B., {Decker}, R.~B., {Roelof}, E.~C., {Krimigis},
  S.~M., {Aran}, A., \& {Sanahuja}, B. 2006, \apj, 653, 1531

\bibitem[{{Lario} {et~al.}(2003){Lario}, {Roelof}, {Decker}, \&
  {Reisenfeld}}]{Lario2003AdSpR..32..579L}
{Lario}, D., {Roelof}, E.~C., {Decker}, R.~B., \& {Reisenfeld}, D.~B. 2003,
  Advances in Space Research, 32, 579

\bibitem[{{Lario} {et~al.}(1998){Lario}, {Sanahuja}, \&
  {Heras}}]{Lario1998ApJ...509..415L}
{Lario}, D., {Sanahuja}, B., \& {Heras}, A.~M. 1998, \apj, 509, 415

\bibitem[{{Lee}(1983)}]{Lee1983JGR....88.6109L}
{Lee}, M.~A. 1983, \jgr, 88, 6109

\bibitem[{{Li} {et~al.}(2003){Li}, {Zank}, \& {Rice}}]{Li2003JGRA..108.1082L}
{Li}, G., {Zank}, G.~P., \& {Rice}, W.~K.~M. 2003, Journal of Geophysical
  Research (Space Physics), 108, 1082

\bibitem[{{Li} {et~al.}(2005){Li}, {Zank}, \& {Rice}}]{Li2005JGRA..110.6104L}
---. 2005, Journal of Geophysical Research (Space Physics), 110, 6104

\bibitem[{{Maclennan} {et~al.}(2001){Maclennan}, {Lanzerotti}, \&
  {Roelof}}]{Maclennan2001AGUSM..SH31A05M}
{Maclennan}, C.~G., {Lanzerotti}, L.~J., \& {Roelof}, E.~C. 2001, AGU Spring
  Meeting Abstracts, 31

\bibitem[{{Matthaeus} {et~al.}(2003){Matthaeus}, {Qin}, {Bieber}, \&
  {Zank}}]{Matthaeus2003ApJ...590L..53M}
{Matthaeus}, W.~H., {Qin}, G., {Bieber}, J.~W., \& {Zank}, G.~P. 2003, \apjl,
  590, L53

\bibitem[{Mazur {et~al.}(2000)Mazur, Mason, Dwyer, Giacalone, Jokipii, \&
  Stone}]{mazur2000ApJInterplanetary}
Mazur, J., Mason, G., Dwyer, J., Giacalone, J., Jokipii, J., \& Stone, E. 2000,
  The Astrophysical Journal Letters, 532, L79

\bibitem[{{McKibben}(1972)}]{McKibben1972JGR....77.3957M}
{McKibben}, R.~B. 1972, \jgr, 77, 3957

\bibitem[{{McKibben} {et~al.}(2001{\natexlab{a}}){McKibben}, {Lopate}, \&
  {Zhang}}]{McKibben2001SSRv}
{McKibben}, R.~B., {Lopate}, C., \& {Zhang}, M. 2001{\natexlab{a}}, \ssr, 97,
  257

\bibitem[{{McKibben} {et~al.}(2001{\natexlab{b}}){McKibben}, {Connell},
  {Lopate}, {Zhang}, {Balogh}, {Marsden}, {Sanderson}, {Tranquille}, {Anglin},
  {Kunow}, {M{\"u}ller-Mellin}, {Heber}, {Raviart}, {Paizis}, \& {COSPIN
  Collaboration}}]{McKibben2001ICRC....8.3281M}
{McKibben}, R.~B., {et~al.} 2001{\natexlab{b}}, in International Cosmic Ray
  Conference, Vol.~8, International Cosmic Ray Conference, 3281--+

\bibitem[{{Ng} {et~al.}(2003){Ng}, {Reames}, \&
  {Tylka}}]{Ng2003ApJ...591..461N}
{Ng}, C.~K., {Reames}, D.~V., \& {Tylka}, A.~J. 2003, \apj, 591, 461

\bibitem[{{Ng} {et~al.}(2012){Ng}, {Reames}, \&
  {Tylka}}]{Ng2012AIPC.1436..212N}
{Ng}, C.~K., {Reames}, D.~V., \& {Tylka}, A.~J. 2012, in American Institute of
  Physics Conference Series, Vol. 1436, American Institute of Physics
  Conference Series, ed. J.~{Heerikhuisen}, G.~{Li}, N.~{Pogorelov}, \&
  G.~{Zank}, 212--218

\bibitem[{{Potgieter} \& {Moraal}(1985)}]{Potgieter1985ApJ...294..425P}
{Potgieter}, M.~S., \& {Moraal}, H. 1985, \apj, 294, 425

\bibitem[{{Qin} {et~al.}(2002{\natexlab{a}}){Qin}, {Matthaeus}, \&
  {Bieber}}]{Qin2002ApJ...578L.117Q}
{Qin}, G., {Matthaeus}, W.~H., \& {Bieber}, J.~W. 2002{\natexlab{a}}, \apjl,
  578, L117

\bibitem[{{Qin} {et~al.}(2002{\natexlab{b}}){Qin}, {Matthaeus}, \&
  {Bieber}}]{Qin2002GeoRLSubdiffusive}
---. 2002{\natexlab{b}}, \grl, 29, 1048

\bibitem[{Qin \& Shalchi(2009)}]{qin2009pitch}
Qin, G., \& Shalchi, A. 2009, The Astrophysical Journal, 707, 61

\bibitem[{{Qin} \& {Shalchi}(2012)}]{Qin2012AdSpRNumerical}
{Qin}, G., \& {Shalchi}, A. 2012, Advances in Space Research, 49, 1643

\bibitem[{Qin \& Shalchi(2014)}]{qin2014detailed}
Qin, G., \& Shalchi, A. 2014, Physics of Plasmas (1994-present), 21, 042906

\bibitem[{Qin {et~al.}(2013)Qin, Wang, Zhang, \& Dalla}]{qin2013transport}
Qin, G., Wang, Y., Zhang, M., \& Dalla, S. 2013, The Astrophysical Journal,
  766, 74

\bibitem[{Qin \& Zhang(2014)}]{qin2014modification}
Qin, G., \& Zhang, L.-H. 2014, The Astrophysical Journal, 787, 12

\bibitem[{Qin {et~al.}(2005)Qin, Zhang, Dwyer, Rassoul, \&
  Mason}]{qin2005model}
Qin, G., Zhang, M., Dwyer, J., Rassoul, H., \& Mason, G. 2005, The
  Astrophysical Journal, 627, 562

\bibitem[{{Qin} {et~al.}(2006){Qin}, {Zhang}, \&
  {Dwyer}}]{Qin2006JGRA..11108101Q}
{Qin}, G., {Zhang}, M., \& {Dwyer}, J.~R. 2006, Journal of Geophysical Research
  (Space Physics), 111, 8101

\bibitem[{{Reames}(1995)}]{Reames1995RvGeS..33..585R}
{Reames}, D.~V. 1995, Reviews of Geophysics Supplement, 33, 585

\bibitem[{{Reames}(1999)}]{Reames1999SSRv90413R}
---. 1999, \ssr, 90, 413

\bibitem[{{Reames}(2010)}]{Reames2010SoPh..265..187R}
---. 2010, \solphys, 265, 187

\bibitem[{{Reames}(2013)}]{Reames2013SSRv..175...53R}
---. 2013, \ssr, 175, 53

\bibitem[{{Reames} {et~al.}(1996){Reames}, {Barbier}, \&
  {Ng}}]{Reames1996ApJ...466..473R}
{Reames}, D.~V., {Barbier}, L.~M., \& {Ng}, C.~K. 1996, \apj, 466, 473

\bibitem[{{Reames} {et~al.}(1997){Reames}, {Kahler}, \&
  {Ng}}]{Reames1997ApJ...491..414R}
{Reames}, D.~V., {Kahler}, S.~W., \& {Ng}, C.~K. 1997, \apj, 491, 414

\bibitem[{{Rice} {et~al.}(2003){Rice}, {Zank}, \&
  {Li}}]{Rice2003JGRA..108.1369R}
{Rice}, W.~K.~M., {Zank}, G.~P., \& {Li}, G. 2003, Journal of Geophysical
  Research (Space Physics), 108, 1369

\bibitem[{{Roelof} {et~al.}(1992){Roelof}, {Gold}, {Simnett}, {Tappin},
  {Armstrong}, \& {Lanzerotti}}]{Roelof1992GeoRL..19.1243R}
{Roelof}, E.~C., {Gold}, R.~E., {Simnett}, G.~M., {Tappin}, S.~J., {Armstrong},
  T.~P., \& {Lanzerotti}, L.~J. 1992, \grl, 19, 1243

\bibitem[{{Schlickeiser}(2002)}]{schlickeiser2002cosmic}
{Schlickeiser}, R. 2002, {Cosmic Ray Astrophysics}, ed. {Schlickeiser, R.}

\bibitem[{{Shalchi} {et~al.}(2004){Shalchi}, {Bieber}, {Matthaeus}, \&
  {Qin}}]{Shalchi2004ApJ...616..617S}
{Shalchi}, A., {Bieber}, J.~W., {Matthaeus}, W.~H., \& {Qin}, G. 2004, \apj,
  616, 617

\bibitem[{{Shalchi} {et~al.}(2010){Shalchi}, {Li}, \&
  {Zank}}]{Shalchi2010Ap&SS.325...99S}
{Shalchi}, A., {Li}, G., \& {Zank}, G.~P. 2010, \apss, 325, 99

\bibitem[{{Skilling}(1971)}]{Skilling1971ApJ...170..265S}
{Skilling}, J. 1971, \apj, 170, 265

\bibitem[{{Sokolov} {et~al.}(2004){Sokolov}, {Roussev}, {Gombosi}, {Lee},
  {K{\'o}ta}, {Forbes}, {Manchester}, \& {Sakai}}]{Sokolov2004ApJ...616L.171S}
{Sokolov}, I.~V., {Roussev}, I.~I., {Gombosi}, T.~I., {Lee}, M.~A., {K{\'o}ta},
  J., {Forbes}, T.~G., {Manchester}, W.~B., \& {Sakai}, J.~I. 2004, \apjl, 616,
  L171

\bibitem[{{Tan} {et~al.}(2009){Tan}, {Reames}, {Ng}, {Saloniemi}, \&
  {Wang}}]{Tan2009ApJ...701.1753T}
{Tan}, L.~C., {Reames}, D.~V., {Ng}, C.~K., {Saloniemi}, O., \& {Wang}, L.
  2009, \apj, 701, 1753

\bibitem[{{Tylka} \& {Lee}(2006)}]{Tylka2006ApJ...646.1319T}
{Tylka}, A.~J., \& {Lee}, M.~A. 2006, \apj, 646, 1319

\bibitem[{{Verkhoglyadova} {et~al.}(2009){Verkhoglyadova}, {Li}, {Zank}, {Hu},
  \& {Mewaldt}}]{Verkhoglyadova2009ApJ...693..894V}
{Verkhoglyadova}, O.~P., {Li}, G., {Zank}, G.~P., {Hu}, Q., \& {Mewaldt}, R.~A.
  2009, \apj, 693, 894

\bibitem[{{Verkhoglyadova} {et~al.}(2010){Verkhoglyadova}, {Li}, {Zank}, {Hu},
  {Cohen}, {Mewaldt}, {Mason}, {Haggerty}, {von Rosenvinge}, \&
  {Looper}}]{Verkhoglyadova2010JGRA..11512103V}
{Verkhoglyadova}, O.~P., {et~al.} 2010, Journal of Geophysical Research (Space
  Physics), 115, 12103

\bibitem[{Wang {et~al.}(2012)Wang, Qin, \& Zhang}]{wang2012effects}
Wang, Y., Qin, G., \& Zhang, M. 2012, The Astrophysical Journal, 752, 37

\bibitem[{{Wang} {et~al.}(2014){Wang}, {Qin}, {Zhang}, \&
  {Dalla}}]{Wang2014ApJ789157W}
{Wang}, Y., {Qin}, G., {Zhang}, M., \& {Dalla}, S. 2014, \apj, 789, 157

\bibitem[{{Zank} {et~al.}(2000){Zank}, {Rice}, \&
  {Wu}}]{Zank2000JGR...10525079Z}
{Zank}, G.~P., {Rice}, W.~K.~M., \& {Wu}, C.~C. 2000, \jgr, 105, 25079

\bibitem[{{Zhang}(1999)}]{Zhang1999ApJ...513..409Z}
{Zhang}, M. 1999, \apj, 513, 409

\bibitem[{{Zhang} {et~al.}(2003){Zhang}, {Jokipii}, \&
  {McKibben}}]{Zhang2003ApJ...595..493Z}
{Zhang}, M., {Jokipii}, J.~R., \& {McKibben}, R.~B. 2003, \apj, 595, 493

\bibitem[{{Zhang} {et~al.}(2009){Zhang}, {Qin}, \&
  {Rassoul}}]{Zhang2009ApJ...692..109Z}
{Zhang}, M., {Qin}, G., \& {Rassoul}, H. 2009, \apj, 692, 109

\bibitem[{{Zuo} {et~al.}(2011){Zuo}, {Zhang}, {Gamayunov}, {Rassoul}, \&
  {Luo}}]{zuo2011}
{Zuo}, P., {Zhang}, M., {Gamayunov}, K., {Rassoul}, H., \& {Luo}, X. 2011,
  \apj, 738, 168

\bibitem[{Zuo {et~al.}(2013)Zuo, Zhang, \& Rassoul}]{zuo2013acceleration}
Zuo, P., Zhang, M., \& Rassoul, H.~K. 2013, The Astrophysical Journal, 767, 6

\end{thebibliography}

\begin{table}
\begin{centering}
\caption {Model Parameters Used in the
Calculations.\label{paratable}}
\begin{tabular} {|l|l|l|} \tableline
Parameter & Physical meaning & Value \\
\tableline\tableline
 $V^{sw}$ & solar wind speed & $400$ km/s\\
\tableline
 $v_s$ & shock speed & $870$ km/s\\
\tableline
 $\phi_s$ & shock width & $60^\circ$\\
\tableline
 $\alpha$ & shock strength  parameter & $2$\tablenotemark{a}\\
\tableline
 $\phi_c$ & shock strength  parameter & $15^\circ$\tablenotemark{b}\\
\tableline
$\gamma$ & injection spectrum &$5.5$\\
\tableline
$B_e$  & magnetic field strength at the Earth & $5$ nT \\
\tableline
$\lambda _\parallel$ & particle radial mean free path & $0.2$ AU\tablenotemark{c}  \\
\tableline
${\bf{\kappa }}_ \bot$ & perpendicular diffusion coeffient  & $0.1 \times {{{\kappa }}_\parallel  }$\tablenotemark{d} \\
\hline
$r_{b0}$ & inner boundary & $0.05$ AU\\
\tableline
$r_{b1}$ & outer boundary & $50$ AU\\
\tableline
\end{tabular}
\end{centering}
\tablenotetext{a}{for $5$ MeV protons.}
\tablenotetext{b}{for $5$ MeV protons.}
\tablenotetext{c}{for $5$ MeV protons in the ecliptic at $1$ AU.}
\tablenotetext{d}{for $5$ MeV protons in the ecliptic at $1$ AU.}

\end{table}

\end{document}